\documentclass[12pt]{iopart}

\usepackage[utf8]{inputenc}
 \expandafter\let\csname equation*\endcsname\relax
 \expandafter\let\csname endequation*\endcsname\relax

\usepackage{iopams,amsmath}
\usepackage{amssymb,latexsym}
\usepackage{amsfonts}
\usepackage{amssymb}
\usepackage{cancel}
\usepackage{subfigure}

\usepackage{graphicx}% Include figure files
\usepackage{bm}
\usepackage{mathrsfs}

\newcommand{\beq}{\begin{equation}}
\newcommand{\eeq}{\end{equation}}	
\newcommand{\beqa}{\begin{eqnarray}}
\newcommand{\eeqa}{\end{eqnarray}}
\newcommand{\beqar}{\begin{eqnarray*}}
\newcommand{\eeqar}{\end{eqnarray*}}

\begin{document}

\title[Sharp bounds on the radius of 
relativistic charged spheres]{Sharp bounds on the radius of 
relativistic charged spheres: Guilfoyle's
stars saturate the Buchdahl-An\-dr\'easson bound}

\author{Jos\'e P. S. Lemos}
%\affiliation 
\address{Centro Multidisciplinar de Astrof\'{\i}sica -- CENTRA,
Departamento de F\'{\i}sica,  Instituto Superior T\'ecnico - IST,
Universidade de Lisboa - UL,
Avenida Rovisco Pais 1, 1049-001 Lisboa, Portugal
\\Email: joselemos@ist.utl.pt}
\author {Vilson T. Zanchin}
%\affiliation 
\address{Centro de Ci\^encias Naturais e Humanas, Universidade
Federal do ABC,  Avenida dos Estados 5001, 09210-580 Santo Andr\'e, S\~ao
Paulo,
Brazil \\ Email: zanchin@ufabc.edu.br}

%\date{today}

\begin{abstract}

Buchdahl, by imposing a few reasonable physical assumptions on the
matter, i.e., its density is a nonincreasing function of the radius
and the fluid is a perfect fluid, and on the configuration, such as
the exterior is the Schwarzschild solution, found that the radius
$r_0$ to mass $m$ ratio of a star would obey the bound $r_0/m\geq9/4$,
the Buchdahl bound.  He also noted that the bound was saturated by the
Schwarzschild interior solution, i.e., the solution with $\rho_{\rm
m}(r)= {\rm constant}$, where $\rho_{\rm m}(r)$ is the energy density
of the matter at $r$, when the central central pressure blows to
infinity.  Generalizations of this bound in various forms have been
studied.  An important generalization was given by Andr\'easson by
including electrically charged matter and imposing a different set of
conditions, namely, $p+2p_T \leq\rho_{\rm m}$, where $p$ is the radial
pressure and $p_T$ the tangential pressure.  His bound is sharp and is
given by $r_0/m\geq9/\left(1+\sqrt{1+3\,q^2/r_0^2}\right)^{2}$, the
Buchdahl-Andr\'easson bound, with $q$ being the total electric charge
of the star.  For $q=0$ one recovers the Buchdahl bound.  However,
following Andr\'easson's proof, the configuration that saturates the
Buchdahl bound is an uncharged shell, rather than the Schwarzschild
interior solution.  By extension, the configurations that saturate the
electrically charged Buchdahl-Andr\'easson bound are charged shells.
One could expect then, in turn, that there should exist an
electrically charged equivalent to the interior Schwarzschild
limit. We find here that this equivalent is provided by the equation
$\rho_{\rm m}(r) + {Q^2(r)}/ {\left(8\pi\,r^4\right)}= {\rm
constant}$, where $Q(r)$ is the electric charge at $r$. This equation
was put forward by Cooperstock and de la Cruz, and Florides, and
realized in Guilfoyle's stars. When the central pressure goes to
infinity Guilfoyle's stars are configurations that also saturate the
Buchdahl-Andr\'easson bound.  It remains to find a proof in Buchdahl's
manner such that these configurations are the limiting configurations
of the bound.

%\pacs{04.70.Bw, 04.20.Jb, 04.40.Nr, 04.40.Dg}

\end{abstract}

% \begin{keyword}
%\keywords
% {Relativistic stars; Schwarzschild interior limit, Buchdahl limit;
% charged stars}
% \end{keyword}

\maketitle

%\end{frontmatter}

\date{today}
\section {Introduction}
\label{sec-introd}

Schwarzschild was the first to put forward the idea of a compact
object \cite{incom_schwarzschild}.  He found an exact solution in
general relativity for a spherical star whose matter content is made
of an incompressible fluid, i.e., $\rho_{\rm m}(r)={\rm constant}$,
where $\rho_{\rm m}(r)$ is the energy density at radius $r$.  This
solution is called the interior Schwarzschild solution.  Moreover he
also established a compactness limit for these configurations, the
interior Schwarzschild limit.  For a star of radius $r_0$ and mass
$m$, the central pressure $p_c$ of the interior Schwarzschild solution
becomes infinite when $r_0/m=9/4$ (we use units such that $G=1$ and
$c=1$, where $G$ is Newton constant of gravitation and $c$ is the
speed of the light.). Thus, for smaller $r_0$ there is certainly no
fluid pressure capable to sustain the configuration against the
gravitational pull, and it will collapse.

The existence of an upper bound for the compactness of the matter
within a star in equilibrium shown by Schwarzschild
\cite{incom_schwarzschild} gave rise to further interest on the
subject. Indeed, Buchdahl \cite{buchdahl} analyzed a spherically
symmetric distribution of matter in a model-independent manner, in
quite general terms.  He imposed a few physically reasonable
restrictions, namely, that the fluid's energy density $\rho_{\rm m}$ is
non-negative and non-increasing outward, the pressure $p$ is
non-negative and isotropic, i.e., the fluid is perfect, and the
boundary defined by $p=0$ is matched to the exterior Schwarzschild
solution. He fount that there was a compactness bound for the
configuration given by $r_0/m\geq9/4$. In his proof it was clear that
the limiting interior Schwarzschild solution (i.e., $\rho={\rm
constant}$ and $p_c\to\infty$) is the configuration that saturates the
bound.

A different proof for Buchdahl's inequality was given by Andr\'easson
\cite{andreq0}. He assumed a different set of restrictions, namely,
that $p+2p_T \leq\rho_{\rm m}$, where $p$ is now the radial pressure,
and $p_T$ is the tangential pressure, with all three quantities being
positive, and the boundary defined by $p=0$ is matched to the exterior
Schwarzschild solution.  With this set of assumptions it was also
found that the compactness bound is Buchdahl's, $r_0/m\geq9/4$.
However, rather than having the limiting interior Schwarzschild
solution as saturating the bound, Andr\'easson found that the bound
was saturated by a compact thin shell.  Indeed, since a thin shell has
by definition $p=0$, a simple exercise shows that for a thin shell
with $2p_T = \rho_{\rm m}$ one finds $r_0/m=9/4$.  Still assuming
$p+2p_T \leq\rho_{\rm m}$ a proof of Buchdahl's bound, different from
\cite{andreq0} was given by Stalker and Karageorgis \cite{sk}.

In order to obtain still lower limits for the bound one has to resort
to matter with extra properties.  The ultimate goal is perhaps to have
a configuration hovering at its own gravitational radius.  One form of
matter that yields repulsion is electrically charged matter. 
Thus, with new compactness limits in mind,
Andr\'easson sought to devise a bound, the Buchdahl-Andr\'easson bound
for electrically charged matter
\cite{andreasson-charged}.  He assumed still that $p+2p_T
\leq\rho_{\rm m}$, and that the boundary is matched to the exterior
Reissner-Nordstr\"om solution.  Relying on Stalker and Karageorgis's
proof \cite{sk}, Andr\'easson found the bound
$r_0/m\geq9/\left(1+\sqrt{1+3\,q^2/r_0^2}\right)^{2}$, where $q$ is
the total electric charge of the star.  
Note that generically all the configurations
that satisfy 
the Buchdahl-Andr\'easson bound are above
the gravitational radius $r_+$ of the star
given by $r_+=m+\sqrt{m^2-q^2}$.
For $q=0$ the Buchdahl bound
$r_0/m\geq9/4$ follows immediately, which
yields an $r_0$ greater than the 
gravitational radius $r_+=2m$.
For $q=r_0$ one obtains
$r_0/m\geq 1$, whose lower bound $r_0=m$ is a configuration at its own
gravitational radius $r_+$, indeed in this extreme 
case $r_0=r_+=m=q$. This latter configuration 
is a quasiblack hole.
Following the previous proof for the uncharged case
\cite{andreq0,sk}, it was found \cite{andreasson-charged}
that the bound is saturated by the
compact thin shells that obey $2p_T = \rho_{\rm m}$.
These shells are such that $r_0/m =
9/\left(1+\sqrt{1+3\,q^2/r_0^2}\right)^{2}$.

A natural question, raised legitimately by Andr\'easson
\cite{andreasson-charged}, is what, if any, are the electrically
charged configurations that saturate the bound and that correspond to
the (uncharged) interior Schwarzschild limit configuration.  In other
words, are there electrically charged configurations, other than thin
shells, that saturate the bound?  And if yes, do they correspond in
the uncharged case to the interior Schwarzschild limit configuration?

Here we are able to answer these questions
in the positive.  Defining $\rho_{\rm m}(r)$
as the energy density of the matter at $r$ and $Q(r)$ as the electric
charge at $r$, we show that the equation of state suggested by
Cooperstock and de la Cruz \cite{cooperstock} and Florides
\cite{florides}, namely, $ \rho_{\rm m}(r) + {Q^2(r)}/
{\left(8\pi\,r^4\right)}= {\rm constant}$, allows to build stars
\cite{guilfoyle,lemoszanchin2010}, Guilfoyle's stars, which for 
$p_c\to\infty$, saturate the
Buchdahl-Andr\'easson bound.  This is not a complete 
surprise.  Cooperstock and
de la Cruz \cite{cooperstock} and Florides \cite{florides} argue
forcefully that the above equation of state is the electric charge
equivalent to the incompressible equation of state set by
Schwarzschild for the interior solution.  Given that these
electrically charged configurations are the analogous to the interior
Schwarzschild configurations, this suggests 
that a proof of the Buchdahl-Andr\'easson bound
along the lines of Buchdahl \cite{buchdahl}, rather than along the
lines of Andr\'easson \cite{andreasson-charged}, should be in sight.
Of course, other electrically charged fluids do not obey
the Buchdahl-Andr\'easson bound. It was shown in 
\cite{lemos1,lemos2} that a fluid with 
$\rho_{\rm m}(r)={\rm contant}$ and with a stiff
electrically charged distribution $Q(r)$, 
different from 
the Cooperstock-de la Cruz-Florides distribution, 
does not saturate the Buchdahl-Andr\'easson bound,
although it comes close to.

The present paper is organized as follows.  In Sec.~\ref{sec-sol} we
write the particular solution of Guilfoyle's work that interests us
here, putting the relevant relations in the appropriate form for
further analysis. Section~\ref{sect-interior} is dedicated to find the
electrical interior Schwarzschild limit of the Guilfoyle solution. We
take the central pressure to infinity and obtain a constraint among
the parameters of the solution.  We show that this constraint
saturates the Buchdahl-Andr\'easson bound. In
Sec.~\ref{sect-conclusion} we conclude.

\section{The Guilfoyle solutions}
\label{sec-sol}

The Guilfoyle star solutions \cite{guilfoyle} are exact
equilibrium solutions of Einstein-Maxwell equations
that were put forward to represent 
electrically 
charged stars.
The metric is conveniently written in the
spherically symmetric form, namely,
\beq
ds^2 = -B(r)\,dt^2 + A(r)\,dr^2 + r^2\left(d\theta^2 +
\sin^2\theta\, d\varphi^2\right) . \label{metricsph}
\eeq
where $t,\, r,\, \theta,\, \varphi,\,$ are the usual Schwarzschild
coordinates, and the metric functions $A(r)$ and $B(r)$ depend on the
radial coordinate $r$ only.

The stars are made of a cold electrically charged fluid with pressure,
have a spherical surface of radius at $r=r_0$, the star's surface,
where the matter pressure vanishes, and exterior to that, for $r>r_0$,
the spacetime is the electromagnetic vacuum, whose metric functions
$A(r)$ and $B(r)$, and electric potential $\phi(r)$,
are given by the
Reissner-Nordstr\"om solution
\beqa
B(r)&=& A^{-1}(r)= 1 -\frac{2m}{r}+\frac{q^2}{r^{2}} \,,
\label{f(r)} \label{RNST}\\
\phi(r)& =& \frac{q}{r} +{\rm const.} , \label{phi-RN}
\eeqa
with $m$ and $q$ being respectively the total mass and total charge of
the stars.  One of the zeros of the metric functions of
equation~(\ref{RNST}) gives the gravitational radius $r_+$, i.e.,
\beq
r_+=m+\sqrt{m^2-q^2}\,.
\label{gravr}
\eeq
The gravitational radius $r_+$ is the horizon radius if the solution
is a vacuum solution.  The other zero of the metric functions of
equation~(\ref{RNST}) is the Cauchy horizon radius $r_-$, given by
$r_-=m-\sqrt{m^2-q^2}$. It will not be mentioned further as it will
not be needed in our analysis.

The systems we are interested here belong to a particular class of the
Guilfoyle  solutions. 
The interior of these solutions is characterized 
by the fact that the metric potential $B(r)$ and the
electric potential $\phi(r)$ are functionally related through a
Weyl-Guilfoyle relation, namely, 
$
B(r)= a \left[-\epsilon\,\phi(r)+b\right]^2$,
with $a$ and $b$ being arbitrary constants and $\epsilon=\pm 1$.
We choose the
solutions for which $b=0$.
Another interesting feature of these charged fluids is 
the relation between the energy density $\rho_{\rm m}$, the pressure $p$,
the charge density $\rho_{\rm e}$, 
and the electromagnetic energy density $\rho_{\rm em}$  
defined by
$
\rho_{\rm em } =  {\left(\nabla_i\phi\right)^2}/{8\pi\,B}
$. 
All these quantities satisfy the constraint 
$
\sqrt{a}\, \rho_{\rm e} = \epsilon\, \left[\rho_{\rm m}+ 3 p+
\left(1-a\right) \rho_{\rm em}\,\right] 
$.

The total electric charge within a sphere of radius $r$,
$Q(r)$, which
follows by integrating the only nonzero component of Maxwell equations,
may be written as
$
Q(r) =  \frac{\phi^\prime (r)\,r^{2}}{\sqrt{B(r) \,A(r)}}\, ,
%\label{qspherical}
$
where the prime denotes the derivative with respect to the radial
coordinate $r$ and without loss of generality
an integration constant was set to zero. 
The class of objects we are interested in here 
obey
$
\epsilon \phi(r) = -\sqrt{{B(r)}/{a}}$.
After such a relation, the electric charge inside a spherical surface of
radius $r$ is now given by $
Q(r) = {-\epsilon\,r^{2}\,B^\prime(r)}/\left({\,2\sqrt{a A(r)}\,
B(r)}\right)$.
The Guilfoyle's ansatz is completed by the assumption that the total energy
density, i.e., 
$\rho_{\rm m}(r) +\rho_{\rm em}(r) = \rho_{\rm m}(r) + {Q^2(r)}/
{\left(8\pi\,r^4\right)}$,  is a constant, 
namely,    
\beqa
\rho_{\rm m}(r) + \frac{Q^2(r)}
{8\pi\,r^4}=\dfrac{3}{R^{2}}\,.
\label{ansa1}
\eeqa
with $R$ being a free parameter of the model. 
Equation (\ref{ansa1}) was devised by Cooperstock and
de la Cruz \cite{cooperstock} and Florides \cite{florides}.
With this, Einstein-Maxwell equations
furnish the metric potentials $A(r)$ and $B(r)$,
\beqa
& & A(r) =\left({1 - \dfrac{r^2}{R^2}}\right)^{-1}\, , \label{A-sol}\\
&& B(r)  = \left[\frac{2-a}{a}\left(  k_0\, \sqrt{1 -
\frac{r^2}{R^2}}-k_1\right)\right]^{2a/(a-2)}
\, , \;\;\;\;\; \label{B-sol}
\eeqa
and the 
electric charge $Q(r)$, and the fluid quantities $\rho_{\rm m}(r)$
and $p(r)$, 
\beqa 
  & & Q(r) = \frac{\epsilon \sqrt{a\,}}{2-a}\frac{k_0\,r^3}{R^2} 
  \left({k_0\, \sqrt{1 - \frac{r^2}{R^2}}-k_1}\right)^{-1}, 
\label{charge-sol1}  \\
 & &8\pi \rho_{\rm m}(r) = \frac{3}{R^2} -\frac{Q^2(r)}{r^4}
	  \, , \label{rhom-sol1} \\
 & & 8\pi p(r) = -\frac{1}{R^2} +  \frac{Q^2(r)} {r^4}   +
\frac{2\varepsilon\, Q(r)}{r^3} \sqrt{1-\dfrac{r^2}{R^2}}\,  ,
\label{p-sol1} \hspace*{.3cm}
\eeqa
with $k_0$ and $k_1$ being integration constants.

There are four constraints.
The continuity of the metric functions on the surface $r=r_0$ yields
two constraints. Another constraint comes from the continuity of the
first derivative of the metric function $B(r)$ across the boundary
surface $r=r_0$, implying that the pressure vanishes at the
boundary. An additional constraint comes from the fact that the
solution must be of Weyl-Guilfoyle type, i.e., 
here satisfying the relation
$
B(r)= a \phi^2(r)$
throughout the spacetime, even at
the boundary. The imposition of
smooth boundary conditions on 
the boundary surface $r=r_0$,
gives the two constraints that 
determine the
two integrations constants $k_0$ and $k_1$, related 
then to the parameters
$m$, $q$, and $R$, and $r_0 $ by 
\beqa
k_0 &=& \frac{R^2}{r_0^2}
\left(\frac{m}{r_0}-\frac{q^2}{r_0^2}\right) \left(1- \dfrac{r_0^2}
{R^2}\right)^{-1/a}\, ,\label{constk}\\
k_1&=& \!\!k_0\,\sqrt{1-\dfrac{r_0^2}{R^2}}\!
\left[1 - \frac{a}{2-a}\!\dfrac{r_0^2}{R^2}
\left(\dfrac{m}{r_0}-\dfrac{q^2}{r_0^2}
\right)^{\!\!-1}\right]. \label{constk_1} \hspace*{.3cm}
\eeqa
The other two constraints
give the following relations
\beqa
&& \frac{m}{r_0}= 
\frac{1}{2}\left(\frac{r_0^2}{R^2} + \frac{q^2}{r_0^2}\right)\,,
\label{mass}\\
&& 
\frac{m}{q} =   \left(1-a\right) \,\frac{q}{r_0} +
\sqrt{a\left[1+\left(a-1\right)\frac{q^2}{r_0^2}\right]}.
\label{guilfmass}
\eeqa
Hence, from the several parameters of the model, $m,\, q,\, a,\,
r_0,\, \mbox{and}\, R$, three of them remain free.

There are additional constraints. We impose 
the stars are not overcharged, which implies
that the pressure is non-negative. This implies also 
\begin{equation}
a\geq 1\,.
\label{nooverc}
\end{equation}
Furthermore we impose there are
no trapped surfaces, i.e., 
\beq
r_0\geq r_+\,,
\label{notrapped}
\eeq
where we allow for the equality,
and $r_+$ is the gravitational 
radius of the configuration given in 
equation~(\ref{gravr}).
Now, by assumption,
the stars are not overcharged
so that $q/m\leq 1$ and thus
\begin{equation}
\frac{r_0}{q}\geq 1\;\,, \quad \frac{r_0}{m}\geq 1\,,
\label{notrapped2}
\end{equation}
are the constraints.

In the following we analyze the infinite central pressure limit of
these compact stars, and obtain configurations that saturate the
Buchdahl-Andréasson bound.

\section{The electrical interior Schwarzschild limit and the
Buchdahl-Andréasson
bound}
\label{sect-interior}

As we know from the case of the interior Schwarzschild solution, the
lower bound of the ratio $r_0/m$ of the Schwarzschild stars is found
by allowing the central pressure to assume indefinitely high values.
Hence, we search for the range of parameters in 
Guilfoyle's solutions for which the central
pressure reaches indefinitely large values.  The pressure $p$ given in
equation~\eqref{p-sol1} behaves for small $r$ as $p(r)\sim
\frac{r^3}{k_0-k_1}$.  Thus, the pressure is in general well behaved at
$r=0$ unless $k_0-k_1=0$. Therefore, the central pressure tends to
infinity, $p_c\to\infty$, when
\beq
k_0=k_1\,. 
\label{kok1}
\eeq
Using relations~\eqref{constk}, \eqref{constk_1}, and \eqref{mass},
equation~\eqref{kok1}
yields
\begin{equation}
\sqrt{1-\dfrac{r_0^2}{R^2}}\!
\left[1 - \frac{2a}{2-a}\dfrac{r_0^2}{R^2}
\left(\dfrac{r_0^2}{R^2}-\dfrac{q^2}{r_0^2}
\right)^{-1}\right]= 1.  \label{pclimit1}
\end{equation}
Solving for $a$ equation~\eqref{guilfmass}, substituting it into the last
equation, and then solving for $q^2/R^2$ we find two solutions.  One
of them yields values of charge whose respective Guilfoyle solutions
are in a range of parameters which do not correspond to charged
stars. Namely, they are black holes with infinite central pressure.
The second solution is the one that has interest in the present context, and
is given by
\begin{equation} 
 \dfrac{q^2}{r_0^2}= 4\left(1-  
\sqrt{ 1-\dfrac{r_0^2}{R^2} } \right)-3\dfrac{r_0^2}{R^2}. 
\label{q2-buchdahl}      
\end{equation}
Equation~\eqref{mass} can be written
for $\frac{r_0^2}{R^2}$, i.e., 
$\frac{r_0^2}{R^2}= 
\frac{2m}{r_0}-\frac{q^2}{r_0^2}$, which 
upon inserting in equation~\eqref{q2-buchdahl} gives the radius to mass
ratio as a function of the electric charge for the electrical interior
Schwarzschild limit, namely,
$
\frac{r_0}{m} = \left[\frac{2}{9}\left(1 + \sqrt{1+
\frac{3q^2}{r_0^2}}\right)+\frac{q^2}{3r_0^2 }\right]^{-1}
%\label{r0to-m-q2}
$,
or equivalently, 
\begin{equation}
 \frac{r_0}{m}=
\frac{1}{\left(\frac{1}{3}+\sqrt{\frac{1}{9}+\frac{q^{2}}{3r_0^{2}
} }
\right)^{2}}\,, 
\label{r0to-m-q2b}
\end{equation}
which is the upper limit of compactness 
for these stars.

Consider, now the sharp bound found by Andr\'easson
\cite{andreasson-charged},
\begin{equation}
 \frac{r_0}{m}\geq
\frac{1}{\left(\frac{1}{3}+\sqrt{\frac{1}{9}+\frac{q^{2}}{3r_0^{2}
} }
\right)^{2}}\,,  
\label{andrea-m2r}
\end{equation}
where $m$ and $q$ are respectively the mass and the charge of the star. 
This constraint is found considering regular
solutions, and the upper bound is met considering infinitely thin shells,
and, moreover, the condition $r_0^2/q^2\geq1$ must be satisfied. 
We see that the limit 
equation~\eqref{r0to-m-q2b} saturates 
the bound equation~\eqref{andrea-m2r},
i.e., the  
electrical interior Schwarzschild (in other words, the infinite pressure) 
limit of the Guilfoyle solution
saturates the Buchdahl-Andréasson bound.

\begin{figure}[ht]
\vskip -.0cm
\begin{center}
\includegraphics[scale=1.1]{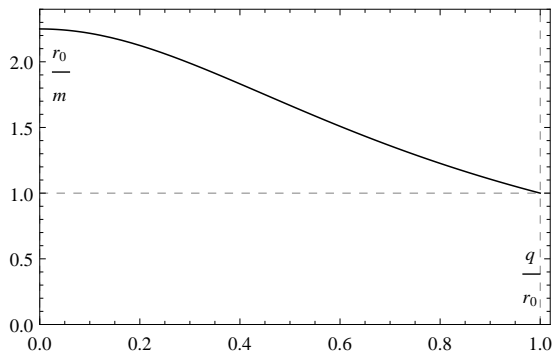}
\caption{The line represents the star radius to mass
ratio as a function of the charge to  star radius
ratio in the electric interior Schwarzschild 
limit.
For each value of the charge ratio $q/r_0$ it
corresponds an $r_0/m$ which is precisely
the electric interior Schwarzschild 
limit for that electric
charge. The limit saturates the Buchdahl-Andr\'easson
bound.
For $q/r_0=0$ one finds the  Schwarzschild result 
$r_0/m =9/4$ that saturates the Buchdahl bound, 
for $q/r_0=1$ one finds 
$r_0/m = 1$, i.e., 
a quasiblack hole configuration.
}
\label{buchdahl-line}
\end{center}
\end{figure}

The
corresponding values for $r_0/m$ of equation~\eqref{r0to-m-q2b} 
as a function of $q/r_0$ are shown  in figure~\ref{buchdahl-line}. 
Note that the limit of zero charge $q/r_0=0$ gives $r_0/m =9/4$, 
the Schwarzschild result for uncharged stars
\cite{incom_schwarzschild}
that saturates the Buchdahl bound \cite{buchdahl}. 
The limit $q/r_0=1$ gives 
$r_0/m = 1$, i.e., 
the quasiblack hole limit, the configuration 
where the star radius $r_0$
touches the horizon radius $r_+$,
$r_0=r_+=m=q$ in this case
\cite{lemoszanchin2010}.

\begin{figure}[ht]
\vskip -.0cm
\begin{center}
\includegraphics[scale=1.1]{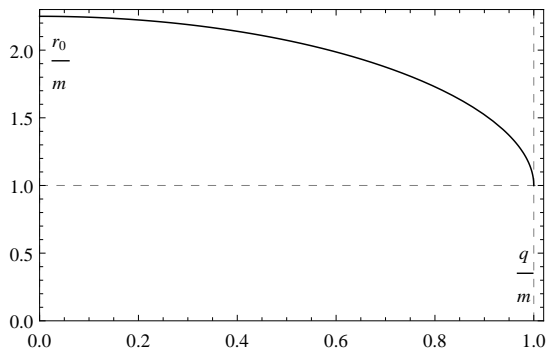}
\caption{
The line represents the star radius to mass
ratio as a function of the charge to star mass 
ratio in the electric interior Schwarzschild 
limit.
For each value of the charge ratio $q/m$ it
corresponds an $r_0/m$ which is precisely
the electric interior Schwarzschild 
limit for that electric
charge. The limit saturates the Buchdahl-Andr\'easson
bound.
For $q/m=0$ one finds the  Schwarzschild result 
$r_0/m =9/4$ that saturates the Buchdahl bound, 
for $q/m=1$ one finds 
$r_0/m = 1$, i.e., 
a quasiblack hole configuration.
}
\label{buchdahl-lineform}
\end{center}
\end{figure}

It would be interesting to find the Buchdahl-Andréasson bound,
equation~\eqref{andrea-m2r}, as a function of $q/m$ instead of $q/r_0$
as it is easier to understand the approach to extremality with the
ratio $q/m$.  Note that the electric Schwarzschild limit,
equation~\eqref{r0to-m-q2b}, can be manipulated to give a cubic
equation for the bound
\begin{equation}
\left(\frac{r_0}{m}\right)^3-
\frac94\left(\frac{r_0}{m}\right)^2+
\frac32\left(\frac{q}{m}\right)^2\left(\frac{r_0}{m}\right)-
\frac14\left(\frac{q}{m}\right)^4\geq0\,.
\label{r0tom-qtomeq}
\end{equation}
Unfortunately this is a nasty cubic equation.  Nevertheless, one can
immediately see from equation ~\eqref{r0tom-qtomeq} that for $q/m=0$
the bound is the Buchdahl bound \cite{incom_schwarzschild},
${r_0}/{m}\geq9/4$, and for the extremal limit $q/m=1$ the bound is
${r_0}/{m}\geq1$, i.e., the quasiblack hole limit
\cite{lemoszanchin2010}.  The solution of
equation~\eqref{r0tom-qtomeq} can be put in a non-illuminating form,
namely, $\frac{r_0}{m}\geq \frac{3}{4} +\frac{1}{4}\left(9
-\frac{8q^2}{m^2}\right)\, {\left[27 - \frac{4q^2}{m^2}\left(9 -
\frac{2q^2}{m^2}- \frac{2q}{m} \sqrt{\frac{q^2}{m^2}-1}
\right)\right]^{-1/3}} + \frac{1}{4}\left[27 - \frac{4q^2}{m^2}\left(9
- \frac{2q^2}{m^2}- \frac{2q}{m} \sqrt{\frac{q^2}{m^2}-1}
\right)\right]^{1/3}$, where since ${\frac{q^2}{m^2}-1}$ is negative
one has to deal with complex numbers.  There are three real solutions,
one should take the branch that interpolates between the values $9/4$
and $1$.  More descriptive, in figure~\ref{buchdahl-lineform} the
bound is displayed for $r_0/m$ as a function of $q/m$.  The Buchdal
\cite{buchdahl} and the quasiblack hole \cite{lemoszanchin2010} limits
stand out clearly.

\begin{figure}[ht]
\vskip -.0cm
\begin{center}
\includegraphics[scale=1.1]{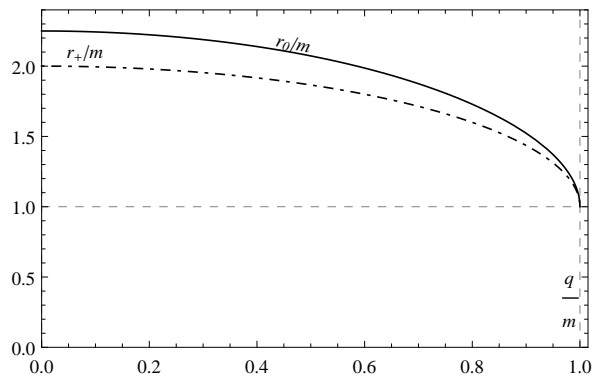}
\caption{
The solid line is the same as in figure~\ref{buchdahl-lineform}.
The dot-dashed line
represents the gravitational radius to mass
ratio as a function of the charge to star mass 
ratio. 
For $q/m=0$ one finds 
$r_0/m=9/4=2.25$ and the  Schwarzschild radius 
$r_+/m = 2$, 
for $q/m=1$ one finds that both radii are equal,  
$r_0/m=1$ and 
$r_+/m=1$, i.e., 
a quasiblack hole configuration.
}
\label{buchdahl_rplus-lines}
\end{center}
\end{figure}

Finally we compare 
the line $r_0/m$ of the Buchdahl-Andr\'easson
bound
as a function of $q/m$ given through equation~(\ref{r0tom-qtomeq})
with the gravitational radius line $r_+/m$
as a function of $q/m$, 
\beq
\frac{r_+}{m}=1+\sqrt{1-\frac{q^2}{m^2}}\,,
\label{gravr2}
\eeq
see equation~(\ref{gravr}).
The comparison of the two lines is done neatly in 
figure~\ref{buchdahl_rplus-lines}.
It is clear from the figure that the
star radius $r_0/m$ is always above 
the star's gravitational radius $r_+/m$, 
except for the quasiblack hole 
configuration where the two radii are
coincident.

\section{Conclusions}
\label{sect-conclusion}

We have shown that in addition to the electrically charged thin shells
with $p+2p_T =\rho_{\rm m}$, the infinite central pressure
Guilfoyle's stars also obey the Buchdahl-Andr\'easson bound,
$r_0/m\geq9/\left(1+\sqrt{1+3\,q^2/r_0^2}\right)^{2}$.  It would be
interesting to find a proof in Buchdahl's manner such that these
infinite central pressure configurations are the limiting
configurations of the bound.

\section*{Acknowledgments}

JPSL thanks the Funda\c c\~ao para a Ci\^encia e a Tecnologia of
Portugal - FCT for support, Project No.~PEst-OE/FIS/UI0099/2014.  VTZ
would like to thank Conselho Nacional de Desenvolvimento Cient\'ifico
e Tecnol\'ogico - CNPq, Brazil, for grants, and Funda\c{c}\~ao de
Amparo \`a Pesquisa do Estado de S\~ao Paulo for a grant, Project No.
2012/08041-5.  JPSL and VTZ thank Coordena\c{c}\~ao de Aperfei\c{c}oamento 
do
Pessoal de N\'\i vel Superior - CAPES, Brazil, for support within the
Programa CSF-PVE, Project No.~88887.068694/2014-00.

\end{document}